# Asymptomatic Phase and Convergent Evolution

# of Coronavirus

J. C. Phillips

Physics and Astronomy, Rutgers University, Piscataway, N J  08854

## Abstract

CoV2019 has evolved to be much more dangerous than CoV2003.   Experiments suggest that structural rearrangements dramatically enhance CoV2019 activity.   We identify a new first stage of infection that precedes structural rearrangements by using biomolecular evolutionary theory to identify sequence differences enhancing viral attachment rates.   We find a small cluster of four single mutations which show that CoV-2 has a new feature that promotes much stronger viral attachment and enhances contagiousness.   The extremely dangerous dynamics of human coronavirus infection is a dramatic example of evolutionary approach of self-organized networks to criticality. It may favor a very successful vaccine.   The identified mutations can be used to test the present theory experimentally.   The theory also works well for the newer strains and explains their increased contagiousness.

## Introduction

Coronaviruses were not considered to be highly pathogenic to humans until the outbreak of severe acute respiratory syndrome (SARS) in 2003 in China[1].   Early discussions of their molecular structure, function and evolution were based on analogies with corresponding functions from flu and other viruses[1,2].   These discussions exhibited local contacts of a few spike receptor binding domain (RBD) amino acids to a few amino acids of respiratory enzymes (ACE2) determined from static crystal structures. Emphasis was placed on similarities and differences between animal and human (CoV-1) viruses. From these static studies alone it is not clear why CoV-2 is so much more contagious than CoV-1, or why several more recent strains containing a few more mutations are even more contagious. Presumably this is caused by convergent evolution[5], which has not occurred recently in other airborne viruses like flu.

Here we review a recent thermodynamic theory of protein dynamics which appears to be especially well suited to discussing the convergent evolution of spikes sequences which contain > 1200 amino acids[3]. Specifically it identifies dynamic features of viral attachment involving specific amino acids responsible for the asymptomatic phase. One might hope to study these with molecular dynamic simulations, but these are limited to proteins with lengths of order 50 amino acids or less, whereas spikes contain ~ 800 amino acids. We conclude that new physics yields important insights into living matter[6,7].

Most physics graduate courses in statistical mechanics end with developments up to 1940. Probably the most important, yet least known, recent developments concern phase transitions in complex networks[8]. Many natural networks exhibit power-law dependence of numbers describing some quality or quantity. The power-law exponents are called fractals, and fractals always characterize second-order phase transitions more accurately than simple fractions[9]. Phase transitions occur because "some biological systems-parts, aspects, or groups of them-may extract important functional benefits from operating at the edge of instability, halfway between order and disorder, i.e., in the vicinity of the critical point of a phase transition. Criticality has been argued to provide biological systems with an optimal balance between robustness against perturbations and flexibility to adapt to changing conditions as well as to confer on them optimal computational capabilities, large dynamical repertoires, unparalleled sensitivity to stimuli, etc. Criticality, with its concomitant scale invariance, can be conjectured to emerge in living systems as the result of adaptive and evolutionary processes"[7].

Proteins are perhaps the most complex networks, so can we reasonably expect to learn much about them from studying simpler complex networks which have a single property described by a single power law and fractal? No, but here there is a hidden simplifying single feature: all proteins have evolved in water. One can learn to quantify protein interactions with water that shape protein transitions from resting states to functional states and return them reversibly to their resting states. This program was begun by biologists in the 1970's, when they began to develop hydropathicity scales. These scales measure how each of the 20 amino acids in proteins "like" (hydrophilic) or "dislike" (hydrophobic) water molecules. The idea proved so popular that by 2000 there were more than 100 biological hydropathicity scales; this has led chemists to conclude that all such scales have only qualitative value[10].

Enter here two Brazilian physicists[11], who realized that the protein structural database had grown enormously by the year 2000. One of the greatest accomplishments of 19th century statistical mechanics was the invention of an equation of state for molecular systems by van der Waals (1873). This equation is used to define a size for each molecule, called its van der Waals radius. The Brazilians revived an old idea, that the hydropathicity of each amino acid in a protein segment could be measured by the area of a centered van der Waals sphere exposed to water. They also added a new idea: that because water folds protein chains into globules, the exposed solvent area (SA) will decrease if one studies protein segments centered on each amino acid. They then examined more than 5000 protein segments. The lengths of their small segments L = 2N + 1 varied from 3 to 45, but the interesting range ran from 9 to 35. Across this range they found linear behavior on a log-log plot for each of the 20 amino acids (aa):

$$\log SA(L) \sim \text{const} - \Psi(aa)\, \text{const.} \log L \qquad (9 \leq L \leq 35)$$

Here $\Psi(aa)$ is recognizable as a network fractal. It arises because the longer segments fold back upon themselves. Thus this striking universal result transcends and compresses thousands of individual protein folding simulations. The 20 fractals based on > 5000 protein segments have effectively compressed a huge amount of three-dimensional structural data into 20 one-dimensional parameters. It was always possible these parameters could have existed in principle, but no one had thought such a simple and sweeping simplification was possible.

We should now pause to digest this remarkable result. Isn't it amazing that the single power laws of other "complex" networks have turned into 20 power laws and 20 fractals for proteins? It appears that each fractal has achieved an accuracy of better than 1%, so the chances that 20 such fractals should occur together, spanning the same segmental length range, are astronomically small. The very special features of proteins as living matter, with each protein using the same menu of amino acids to function through millions of reversible phase transitions before failing, are what have made this happen.

Before going further, one should pause again. Undergraduate thermodynamics teaches us that there are two kinds of phase transitions, called first- and second-order. The second-order transitions involve many long-range interactions, and are much more complex; the Brazilians found a way to quantify them in proteins. Some proteins, especially those involved in metabolic shot-range oxygen redux interactions, are better described by a hydropathic scale based on a first-order effect, such as unfolding from water to air. This problem had been solved by biologists in the old days[12], and although its applications are more limited, it furnishes a necessary check on any hydropathic analysis.

The next step is to specialize our analysis to connect to the convergent evolution of a given protein and its functions. We know a lot about protein functions, and what the structures look like, for instance, in some static binding states. Still, experiments provide few details about dynamical interactions, which can, and usually do, involve the entire protein, not just the binding regions[1,2]. These long-range (or allosteric) interactions have been mysterious and resisted analysis. We do expect that convergent evolution will improve protein functions, probably most clearly in dynamical interactions which shorten functioning times. Proteins often function by rotating some domains from the resting to the functional state and back. It could be optimal for these domains to have similar lengths (like the subcubes in Rubik's cube), which we call W.

Because proteins have been shaped by water, we now plot the hydropathic parameters $\Psi(aa)$ across the entire protein sequences always easily obtained (for instance, from Uniprot). At first, this looks very confusing, because there are hundreds of oscillations. Then we remember the domains and their motion, so we construct a matrix $\Psi(aa,W)$, with entries averaged over sliding windows of width W centered on each amino acid. This has a smoothing effect, and the number of oscillations decreases, as well as their width. The overall width of oscillations of $\Psi(aa,W)$ is calculated using a standard mathematical tool, the variance {(mean of squares) – (square of mean)}.

We are now ready to use convergent evolution to our advantage. We can compare proteins from earlier and later species by taking the ratio of their variances for each W. Features of the resulting plots are often sufficient to suggest optimized values of W. We can plot the profiles Ψ(aa,W) for two or more species, and physicists who have seen features in optical of simple inorganic solids often recognize features in such protein profiles. Profiles of hen egg white, and myoglobin, hemoglobin, and neuroglobin, as well as molecular motors, have evolved in readily interpreted ways[13-15]. Each protein family has a fascinating history.

To see how these general considerations work out in practice, let us take a look at what hen egg white looks like under our evolutionary microscope[16]. Lysozyme *c* (aka Hen Egg White, or HEW) was for some time the most studied protein: the PDB contains more than 200 human and 400 chicken HEW structures. HEW is also present in many other species, not only in the 400 million year old chicken sequence, but in most other vertebrates, almost unchanged in its peptide backbone structure. The backbone structure is exceptionally stable, with human and chicken amino acid binding positions superposable to 1.5A°, while the aa sequence mutates from chicken to human with 60 mutations of its 148 amino acids.

HEW has a nearly centrosymmetric tripartite α helices (1- 56 and 104-148) and β strands (57-103) secondary structure, which is conserved for terrestrial animals. During its long career, HEW has performed at least three functions, as an enzyme, an antibiotic, and most recently an amyloidosis aggregation suppressor. The relative importance of these functions has changed from short-lived to long-lived species, and it seems likely that these changes are reflected in the amino acid mutations that have reshaped and while still maintaining the centrosymmetric structure. In Fig. 1 we see how these changes look when we change the length W of the sliding window of our evolutionary microscope using three different scales[16]. This rich structure contains a lot of information. Its structure is centosymmetrical both hydropathically and elastically, because β strands are elastically stiffer than α helices. The purely elastic features can be monitored by a third scale which measures the inside/outside (or exposed) tendencies of amino acids in β strands[17].

The evolutionary differences in Fig. 1 are all small near W = 1, which is the length scale L. The value L = 1 is used in BLAST, the standard tool often used for starting to compare protein sequences[3]. The MZ scale[11] shows the largest differences, while the KD scale[12] shows the smallest differences. This means that evolution has made the largest shaping tendencies in long-range inter-domain interactions, with smaller changes in short-range intradomain interactions.

The largest MZ peak occurs at W = 69, roughly half the protein length, and reflects overall smoothing of the human shape by evolution. The secondary peaks near 37 and 51 reflect the overall smoothing of the central β strand region. This is clear from plotting the actual Ψ(aa,69) profiles[16]. There is also some MZ smoothing of the C terminal α helices, which is why these peaks are larger than those found with the β strand scale. It seems very likely that this smoothing is enhanced in longer-lived species.

Given a protein hydropathic profile Ψ(aa,W), one notices that it has two kinds of extrema, hydrophobic maxima and hydrophilic minima. It is natural to suppose that the inside maxima act as pivots or pinning points for the conformational domain motion that is functionally significant, while the outer minima act as hinges. This language does not specify the conformational motion in Euclidean space that is functionally significant, as it jumps directly from the universal sequence evolution of Ψ(aa) to function. In practice this method has proved to be better than going elastometrically only between sequence and structure, by using the isotropic vibrational amplitudes of individual amino acids measured in structural studies. The latter cannot be converted into long-range strain fields without knowing distant interatomic force fields in great detail (not possible for large proteins). Again this shows that great simplifications are possible using water interactions just because they are weak, and dominate protein shapes in the length range $9 \le L \le 35$.

What happens when the extrema of an optimized Ψ(aa,W) are nearly level? Examples of such leveling were apparent in studies of several proteins, such as aspirin, where leveling was medically important[18], and this appeared to be a new idea. It is called synchronization, and there are examples of it both in simple mathematical models[19,20] and in many single-fractal networks[8]. This concept is the key to using Ψ(aa,W) profiles to identify the mutations responsible for the asymptomatic phase of Cov19. In practice, with W as the only parameter, the evolutionary convergence of multiple extrema toward synchronization can be unambiguous.

**Results**

The range of protein structures and functions is very large, and thermodynamic methods did not work well for flu viruses, whose strains are variable, but are usually very contagious only when vaccines fail. However, a glance at Coronavirus structure shows that their spikes are long and are immersed in water, which makes the mutational differences between the CoV-1 and CoV-2 amino acid sequences ideal subjects for analysis by hydropathic scaling. The current coronavirus belongs to a virus family that also causes common colds. CoV-1 and CoV-2 are much more dangerous. Worse still, many numbers now show that CoV-2 (2019) spikes have evolved to be much more dangerous than CoV-1(2003) spikes. Much of this "nearly perfect" 2019 viral action[20] can be explained in terms of two well-studied cleavage-trimeric reassembly structural differences of coronal spikes[3]. The spikes contain $> 1200$ amino acids, and BLAST shows that $\sim 300$ of these are mutated. Can only a few of these 300 mutations, far from the two cleavage sites, also be contributing to the extremely strong viral interaction of CoV-2? How do we identify these possibly critical distal sites, which are so far unrecognized in static studies?

Patient readers will have guessed by now that thermodynamic hydropathic scaling is a method that is perfectly suited to answering the obvious questions[21] about the convergent evolution of CoV to becoming the most contagious and persistent virus in history. Schrodinger in his Dublin lectures (1943) asked, What is life?[22] General ideas that have been discussed by physicists (especially since the discovery of DNA) are extended by scaling to include quantitative details of protein evolution, and now lead us to explicit models[23,24] of the asymptomatic stage of CoV-2, now much more prominent than it was in CoV-1.

The overall profiles of CoV-1 and CoV-2 in Fig. 2 seem at first to be very similar, in spite of the 300 mutations in CoV-2. The strong hydrophobic peak at the C terminal end is little changed, but a closer look (Fig. 3) in the central region shows large changes (Table 1). The CoV-1 edge near 550 has shifted upwards (hydrophobically) in CoV-2. . The striking change in minimum 2 is caused by a cluster of four key mutations from CoV-1 to CoV-2. These are (CoV-1 site numbering from Uniprot P59594): 546Gln to Leu; 556 and 561Ser to Ala; and 568Ser to Leu. The differences associated with each of these key mutations are hydropathically large (~50-100 in the MZ scale[24]; all 20 amino acids span a range from most hydrophilic to most hydrophobic of 170). It is this cluster of four single site mutations that bring minimum 2 from CoV-1 into synchronized alignment with minima 1,4, and 5.

Hydropathic synchronization explains the appearance of the asymptomatic phase, because it is presumably associated with viral concentration in the upper respiratory tract, which is moister, and from which aerosol-borne viruses can be expelled. The second stage of infection involves viral fusing with cells through trimeric spike formation[3]. Several new strains have appeared, with further synchronization. Thus the UK strain is about twice as level as CoV-2 (Table 1). At this point the synchronized effects are near the limits of accuracy of the MZ scale. To see how useful the concept of phase transitions is, the profiles of CoV-1 and CoV-2 were compared wit the first-order KD scale[12,24], and there was no edge alignment[23]. Those who are interested in further checks on the unique accuracy of the MZ scale might use any of the other 100+ hydropathicity scales available in the biological literature.attached

A more interesting test of the thermodynamic theory is provided by the new California strain, which has only one additional mutation, L452R[25]. Its effect on edge 1 (centered at 454) can be calculated from the MZ values of L and R, with W = 35. It shifts this edge lower by 3.4 to 133.7. The four edges remain well synchronized, while their overall average has been lowered. Thus the edges are more hydrophilic. This could lead to more attachment in the throat (faster spreading), as well as faster trimeric reassembly in the lungs[16].

**Conclusions**

The level hydrophilic extrema of $\Psi(aa,35)$ of the spikes of CoV-2 may be unique among viruses. These level extrema bring CoV-2 contagion very close to a critical point. Because CoV-2 is very close to a critical point, its symptoms can vary widely between individuals. One of the symptoms is loss of smell and taste. A large statistical survey reported that this was often an "all-or-nothing" effect[26] characteristic of a critical point. Because of interactions with other proteins, the individual reaction can bifurcate on either side of the critical point. Thus very small differences in many genes can cause large fluctuations in infection levels and fatalities between different countries. A recent review concludes that these differences are indeed large worldwide and could have many causes[27]. It also remarks that "epidemiology isn't physics", but the physics of phase transitions provides criticality as a single cause for diverse effects. Another example is the fractal behavior of COV infection curves[28]. Phase transitions even predicted in the abstract of an early paper the high success rate of spike-based vaccines[23].

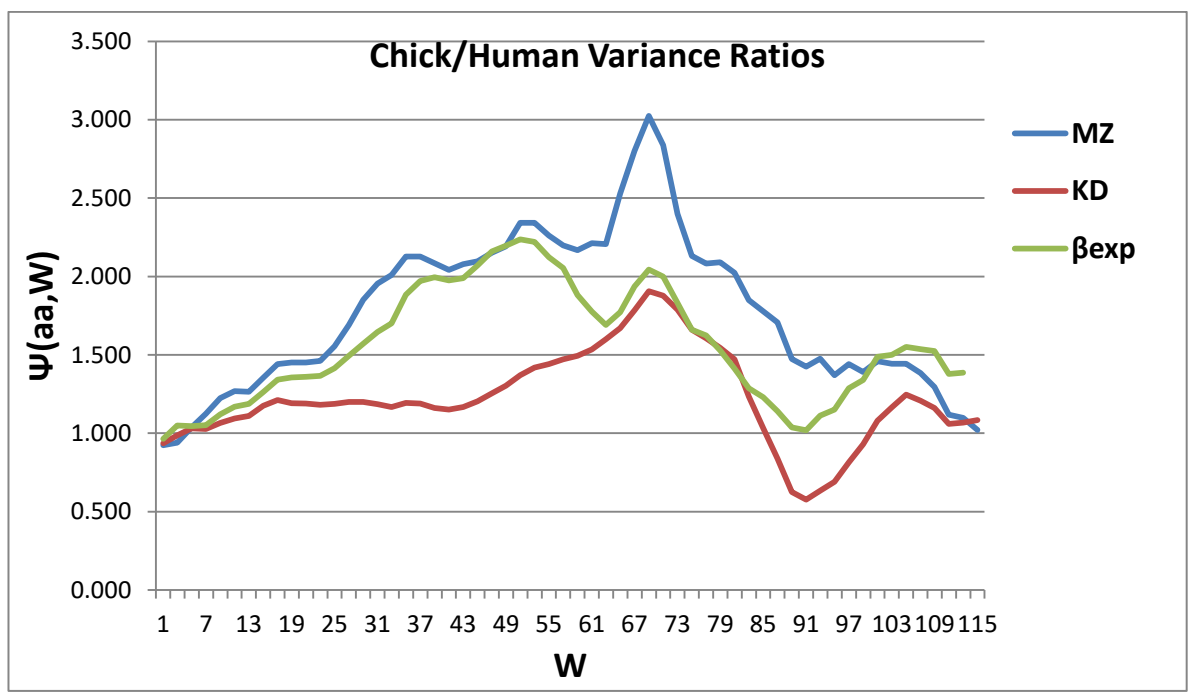

Fig.1   The evolution from chicken to human of the variance ratios of hen egg white

with three different in/out amino acid scales. Values  > 1 mean human is smoother.

Note that near W = 1 the variance ratio differences are small.  The scales describe

(first,second) (MZ,KD)-order hydropathic effects and outer preferences of β strands

(β exposed).

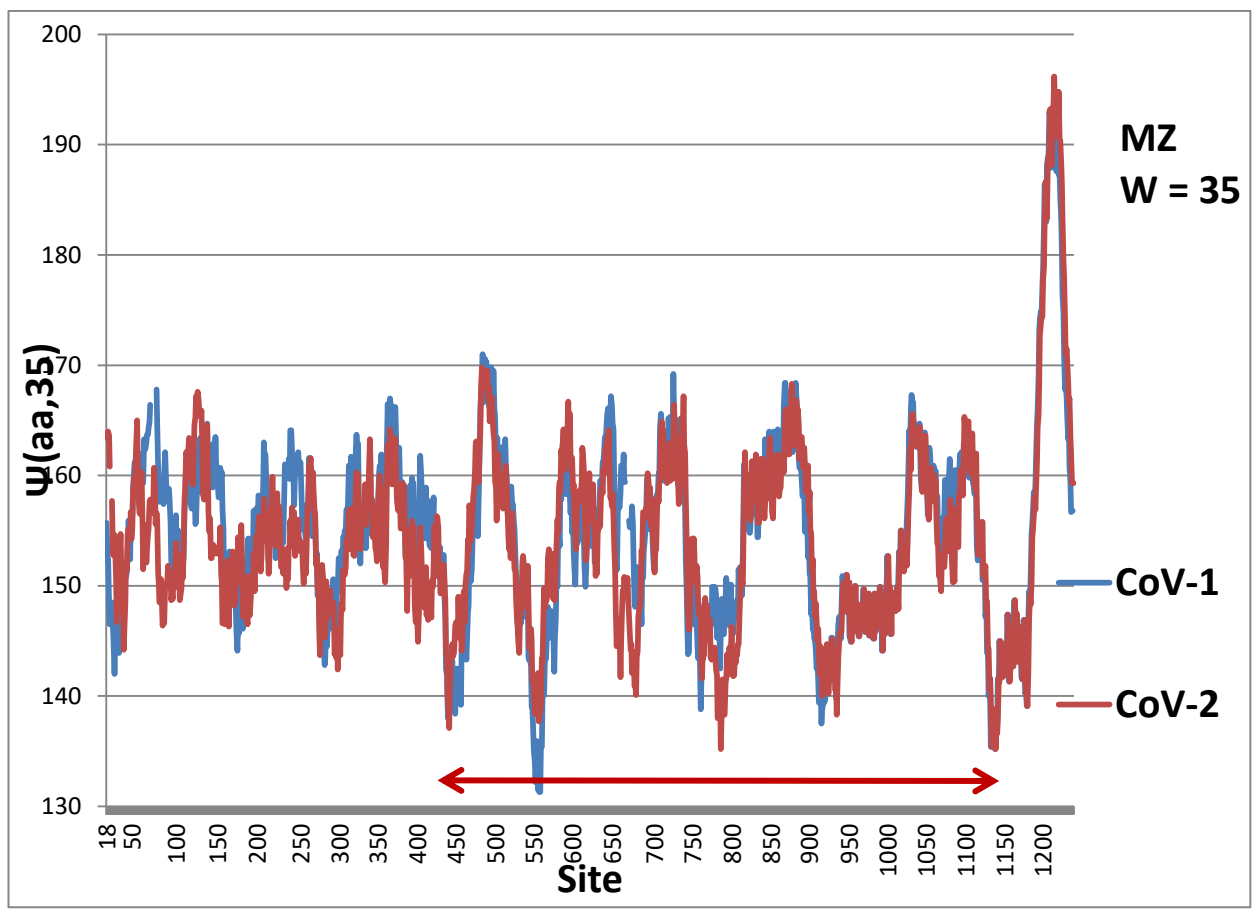

Fig.2.  The hydroprofiles of CoV-1 and CoV-2 are quite similar, but looking closely,

we see a level set appearing CoV-2 near 140, because the CoV-1 edge near 550 has shifted upwards (hydrophobically) in CoV-2.

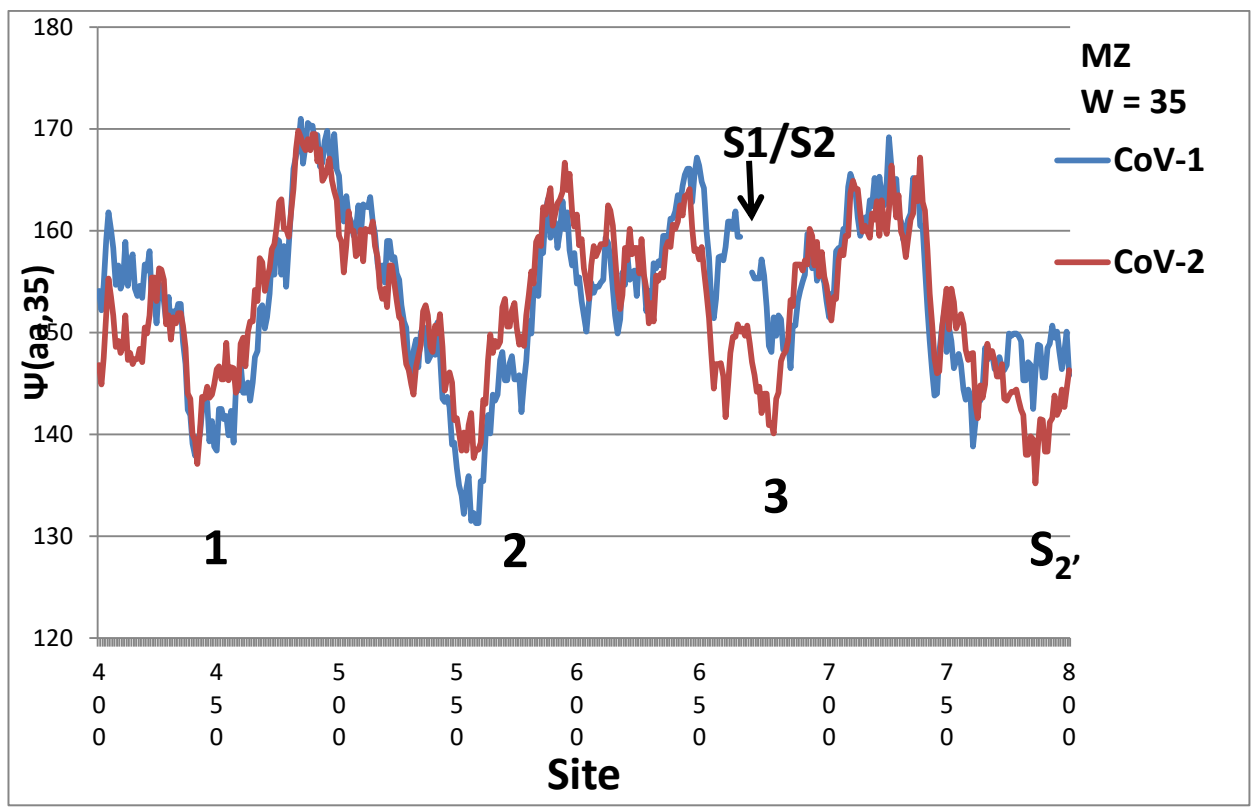

Fig. 3. Enlargement of the central region, including the S1/S2 cleavage sites, shows how the new insertion[3] of PRRA in CoV-2 lowers $\Psi(aa,35)$ to 140.9 at the 3 minimum, aligning it with ~140 at minimum 1. Meanwhile a cluster of four hydrophobic single mutations have raised the very deep minimum of CoV-1 at 559 from near 130 to near 140. Together with the edge near 1150 (Fig. 2) these three edges are all nearly aligned within ~ 1%.

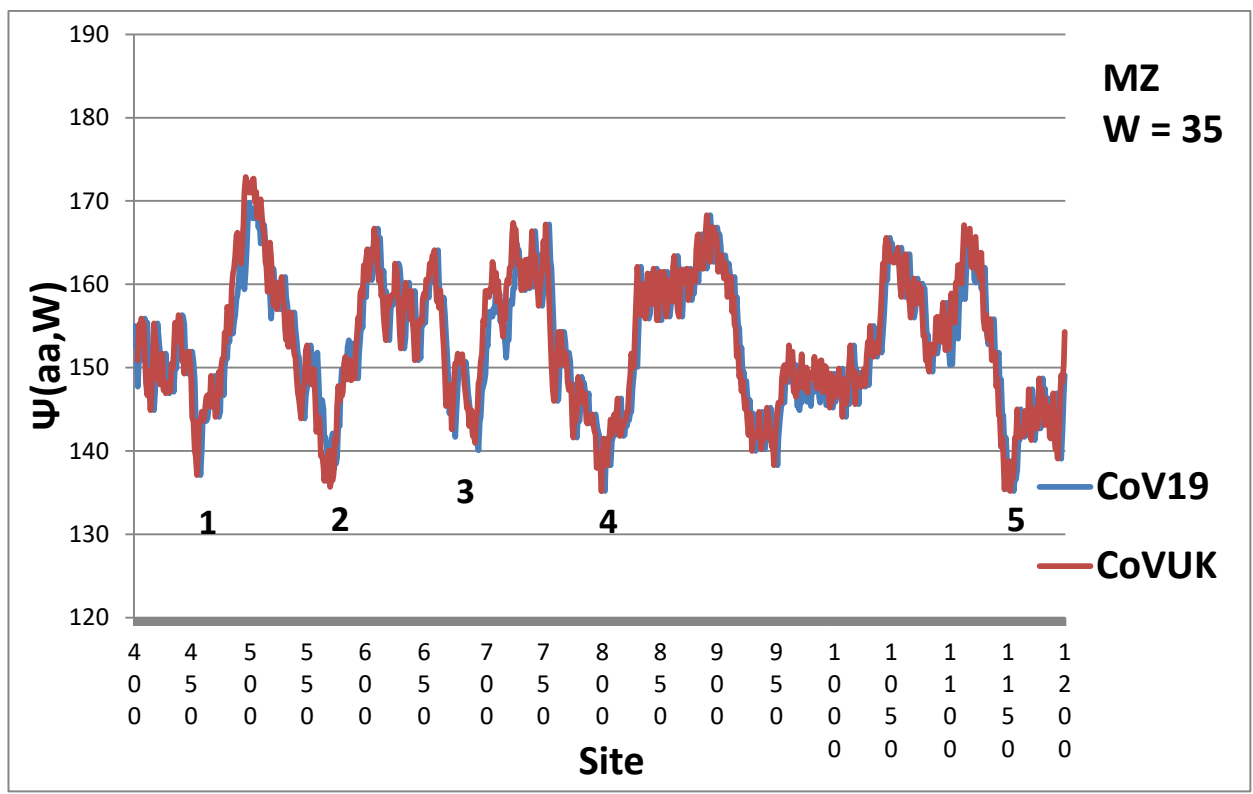

Fig. 4.  There are four nearly level hydrophilic edges in CoV19, numbered 1,2,4 and 5.  When their positions are examined with three-figure accuracy, one finds that the (1,2,4,5) hydrophilic edges of the UK strain are about twice as level as those in CoV19.

| Strain \ Edges | 1 (454) | 2 (569) | 3 (694) | 4 (803) | 5 (1156) |
|---|---|---|---|---|---|
| CoV-1 | 137.9 | 131.3 | 146.5 | 142.5 | 137.5 |
| CoV-2 | 137.1 | 137.7 | 140.1 | 135.2 | 135.2 |
| B.1.1.7 | 137.1 | 135.7 | 141.0 | 135.2 | 135.2 |

Table 1. CoV-1 edges are poorly aligned[23]. For edges (1,2,4,5) the CoV-2 and CoVUK (B.1.1.7) deviations are 4% and 2% of the overall range of $\Psi(R,35)$. These very small deviations are the result of dynamical synchronization[23]. Edge 3 is located near the S1/S2 cleavage site. Note that new mutations can be significant if they lie within 17 amino acids of these five, and involve large (non-synonomous) changes in $\Psi(aa)$. An example of a nonsynonomous change is A to D (change 70), while the synonomous E to K change is only 8[24]. Note that these four edge (1,2,4,5) W = 35 windows occupy only 10% of the viral spike length. It seems very unlikely that four large hydropathic shifts could cluster in exactly the narrow range around edge 2 to make it level with edges 1, 4 (which also shifted into better alignment) and 5. This "unlikely" effect is the result of convergent evolution.